# Fortschritte der Physik

www.fp-journal.org

# Progress of Physics

# Interaction of Gravitational Waves with Superconductors

NA Inan[1,*], JJ Thompson[1], and RY Chiao[2]

Applying the Helmholtz Decomposition theorem to linearized General Relativity leads to a gauge-invariant formulation where the transverse-traceless part of the metric perturbation describes gravitational waves in matter. Gravitational waves incident on a superconductor can be described by a linear London-like constituent equation characterized by a "gravitational shear modulus" and a corresponding plasma frequency and penetration depth. Electric-like and magnetic-like gravitational tensor fields are defined in terms of the strain field of a gravitational wave. It is shown that in the DC limit, the magnetic-like tensor field is expelled from the superconductor in a gravitational Meissner-like effect. The Cooper pair density is described by the Ginzburg-Landau free energy density embedded in curved spacetime. The ionic lattice is modeled by quantum harmonic oscillators coupled to gravitational waves and characterized by quasi-energy eigenvalues for the phonon modes. The formulation predicts the possibility of a dynamical Casimir effect since the zero-point energy of the ionic lattice phonons is found to be modulated by the gravitational wave, in a quantum analog of a "Weber-bar effect." Applying periodic thermodynamics and the Debye model in the low-temperature limit leads to a free energy density for the ionic lattice. Lastly, we relate the gravitational strain of space to the strain of matter to show that the response to a gravitational wave is far less for the Cooper pair density than for the ionic lattice. This predicts a charge separation effect in the superconductor as a result of the gravitational wave.

## 1 Introduction

The fundamental motivation for this paper is to examine how gravity interacts with quantum matter, yielding experimentally testable predictions. In particular, we describe a mechanism which quantifies how gravitational waves interact with superconductors, and may even be expelled by them in a gravitational Meissner-like effect. In addition, we investigate a "charge separation effect," also dubbed the "Heisenberg-Coulomb effect," which was first proposed in [1], and later referenced in [2]. This effect is a result of the Cooper pairs being condensed into a quantum mechanical zero-momentum eigenstate, and therefore, by the Heisenberg Uncertainty Principle, being *non-local*. Hence they cannot follow any classical trajectories in the presence of a gravitational wave.

In contrast to this, each ion in its lattice site can respond *locally* to the gravitational wave, and thereby the ionic lattice can oscillate in its deformation accordingly. It is found that the zero-point energy of the phonon modes of the lattice dominates the response of the lattice to the wave. The *difference* in motion of the Cooper pairs (negatively charged), and the lattice ions (positively charged) results in the charge separation effect. According to [1], this creates an electric field which opposes the supercurrents induced by the gravitational wave. The resulting Coulomb force acts as a very strong restoring force that makes the superconductor extremely "stiff" to gravitational waves. The mass supercurrents produced by this effect will radiate gravitationally. Therefore, the incoming gravitational wave will be re-radiated back out, and the superconductor will act as a mirror to gravitational waves.

It is well established in General Relativity, that gravitational waves can be described by the metric perturbation, $h_{ij}^{TT}$, in the transverse-traceless (TT) gauge. This gauge eliminates unphysical degrees of freedom from the metric leaving only the "plus" and "cross" polarization wave fields, $h_\oplus$ and $h_\otimes$, respectively. [3] However, the TT gauge can only describe plane-waves *in vacuum*. It does not provide a description of gravitational waves *in matter*.

* NA Inan   E-mail: ninan@ucmerced.edu
[1] University of California, Merced, School of Natural Sciences, P.O. Box 2039, Merced, CA 95344, USA
[2] University of California, Merced, Schools of Natural Sciences and Engineering, P.O. Box 2039, Merced, CA 95344, USA

Therefore we begin with a metric which is constructed in accordance with the Helmholtz Decomposition theorem as shown in [4]. This formulation is gauge invariant (to linear order in the metric) and isolates the radiative degrees of freedom which are characterized by the transverse-traceless part[1] of the metric perturbation, $h_{ij}^{\tau\tau}$. Such a formulation is well suited for describing the interaction of gravitational waves with quantum matter such as a superconductor.

## 2 A gravitational London-like constituent equation

Here we motivate a constituent equation for gravitational waves interacting with a superconductor by considering an analogy with the London constituent equation in electromagnetism. The London equation for superconductors is known to be $\mathbf{J}_s = -\Lambda_L \mathbf{A}$, where the supercurrent $\mathbf{J}_s$ is induced by the vector potential $\mathbf{A}$. Here the London constant is known to be $\Lambda_L = n_s e^2/m_e$, with $n_s$ being the number density of Cooper pairs, $e$ the electron charge, and $m_e$ the electron mass. The negative sign in this relationship implies that electromagnetic fields are *expelled* from a superconductor.

For the case of a gravitational wave incident on a superconductor, we also develop a linear relationship between $h_{ij}^{\tau\tau}$, the transverse-traceless strain field for an incident gravitational wave, and $T_{ij}^{\tau\tau}$, the transverse-traceless stress induced by the wave[2]. The constituent equation is

$$T_{ij}^{\tau\tau} = -\mu_G h_{ij}^{\tau\tau} \tag{1}$$

where $\mu_G$ is a positive, real constant with the dimensions of energy density. It can be thought of as a *gravitational shear modulus* which determines how much stress, $T_{ij}^{\tau\tau}$, is produced within a superconductor by an incident gravitational wave field, $h_{ij}^{\tau\tau}$. Similar to the case with the London equation in electromagnetism, the relationship (1) follows as a direct result of the fact that particles within a superconductor (namely, Cooper pairs and lattice ions) undergo *dissipationless* acceleration due to the gravitational wave field. This dissipationless nature follows from the fact that there exists a BCS energy gap that prevents dissipation by low-lying energy states nearby to the BCS ground state. Again, similar to the London equation, the negative sign relating $T_{ij}^{\tau\tau}$ and $h_{ij}^{\tau\tau}$ would imply gravitational waves are *expelled* from a superconductor in a gravitational Meissner-like effect. It will be shown in the following section that this Meissner-like expulsion leads to an associated plasma frequency and penetration depth for gravitational waves incident on a superconductor.

## 3 A gravitational wave plasma frequency and penetration depth

It is shown in [4] that the linearized Einstein equation leads to a wave equation for the transverse-traceless strain field given by[3]

$$\Box h_{ij}^{\tau\tau} = -2\kappa T_{ij}^{\tau\tau} \tag{2}$$

where $\kappa = 8\pi G/c^4$. We can insert (1) into (2), and consider a monochromatic plane wave propagating in the $z$-direction given by $h_{ij}^{\tau\tau}(z,t) = A_{ij}^{\tau\tau} e^{i(kz-\omega t)}$ where $A_{ij}^{\tau\tau}$ is a constant amplitude tensor. This leads to a dispersion relation given by

$$k^2 = \frac{\omega^2}{c^2}\left(1 - \frac{16\pi G \mu_G/c^2}{\omega^2}\right) \tag{3}$$

It is pointed out in [5] that (3) resembles the electromagnetic equation for a dense plasma with a gravitational plasma frequency defined as

$$\omega_G^2 \equiv 16\pi G \mu_G/c^2 \tag{4}$$

We can define a complex wave number as $k = K + i\alpha$, where $K$ and $\alpha$ are real quantities, and insert this into the plane wave solution. Separating the real and imaginary parts of the phase gives $h_{ij}^{\tau\tau}(z,t) = A_{ij}^{\tau\tau} e^{-\alpha z} e^{i(Kz-\omega t)}$. Here we see that the wave falls off exponentially with distance, where $\alpha$ is the exponential decay factor. The square of the wave number is $k^2 = K^2 - \alpha^2 + 2iK\alpha$. Since $k^2$ in (3) is only real, then we must have either $K = 0$ or $\alpha = 0$. For $K = 0$, we use (3) to solve for $\alpha$ and define a frequency-dependent penetration depth as $\delta_G \equiv 1/\alpha$. This gives the following solution.

$$h_{ij}^{\tau\tau}(z,t) = A_{ij}^{\tau\tau} e^{-z/\delta_G} e^{-i\omega t} \tag{5}$$

---

[1] Note that [4] uses $h_{ij}^{TT}$ instead of $h_{ij}^{\tau\tau}$. However, we use $h_{ij}^{TT}$ to refer to the metric perturbation in the transverse-traceless *gauge*. This is not to be confused with $h_{ij}^{\tau\tau}$ which is the transverse-traceless *part* of the metric perturbation.

[2] Note that [4] uses $\sigma_{ij}$ instead of $T_{ij}^{\tau\tau}$.

[3] Note that this formulation is applicable for *all* frequency ranges. The only limitation is that it applies only in the weak-field limit of linearized GR with the requirement that $h_{\mu\nu} \to 0$ as $r \to \infty$.

where

$$\delta_G^2 = \frac{c^2}{16\pi G \mu_G/c^2 - \omega^2} \qquad (6)$$

An exponential decay solution implies that gravitational waves are *expelled* from the superconductor in a gravitational Meissner-like effect, where $\delta_G$ characterizes the depth that gravitational waves penetrate the superconductor. Note the critical role of $\mu_G$ in determining the value of this depth. In later sections, we will formally develop the constituent equation (1) and obtain an expression for $\mu_G$ for the Cooper pairs and the lattice ions.

## 4 A Meissner-like effect for gravitational waves in the DC limit

Recall that using the electromagnetic London constituent equation, $\mathbf{J} = -\Lambda_L \mathbf{A}$, and taking temporal and spatial derivatives, leads to equations involving the electric and magnetic fields within a superconductor given as

$$\partial_t \mathbf{J} = \Lambda_L \mathbf{E} \qquad \text{and} \qquad \nabla \times \mathbf{J} = -\Lambda_L \mathbf{B} \qquad (7)$$

For a sinusoidal current density, we have $\partial_t \mathbf{J} \propto \omega \mathbf{J}$. Therefore, in the DC limit ($\omega \to 0$), the first equation above requires that $\mathbf{E} \to 0$. This implies that in the DC limit, the electric field vanishes completely throughout the entire superconductor and only a magnetic field remains. It can also be shown that the magnetic field satisfies a Yukawa-like equation given as

$$\nabla^2 \mathbf{B} - \frac{1}{\lambda_L^2} \mathbf{B} = 0 \qquad (8)$$

The solution to this equation is $B(z) = B_0 e^{-z/\lambda_L}$ where $z$ is the distance from the surface to the interior of the superconductor, and $\lambda_L$ is the London penetration depth. This result implies that the magnetic field is expelled from the interior of the superconductor which is referred to as the Meissner effect.

Here we find an analogous Meissner-like effect for the DC limit of gravitational waves in a superconductor. First note that for the case of gravitational waves in the far field,[4] the geodesic equation of motion (to first order in the metric perturbation and second order in velocity) is

$$a^i = -\dot{h}_{ij}^{TT} v^j + \left(\tfrac{1}{2}\partial_i h_{jk}^{TT} - \partial_j h_{ik}^{TT}\right) v^j v^k \qquad (9)$$

Recall that in electromagnetism, the propagating degrees of freedom are $\mathbf{E} = -\partial_t \mathbf{A}$ and $\mathbf{B} = \nabla \times \mathbf{A}$ which are the temporal and spatial derivatives of the vector potential, respectively. By analogy, in (9) we observe that the temporal and spatial derivatives of $h_{ij}^{TT}$ appear. Thus, we can define propagating electric-like and magnetic-like *tensor* fields, respectively, as

$$\mathscr{E}_{ij} \equiv -\dot{h}_{ij}^{TT} \qquad \text{and} \qquad \mathscr{B}_{ijk} \equiv \partial_k h_{ij}^{TT} \qquad (10)$$

Using these, we can write the geodesic equation in (9) as

$$a^i = v^j \mathscr{E}_{ij} + v^j v^k \left(\tfrac{1}{2}\mathscr{B}_{jki} - \mathscr{B}_{ikj}\right) \qquad (11)$$

Hence we find that the electric-like and magnetic-like tensor fields defined in (10) determine the physical motion of test particles in the presence of gravitational waves.[5] Next we will use these definitions to identify a Meissner-like effect for gravitational waves in the DC limit. Taking temporal and spatial derivatives of the London-like gravitational constituent equation in (1), and using the definitions in (10), leads to constituent equations given as

$$\partial_t T_{ij}^{TT} = \mu_G \mathscr{E}_{ij} \qquad \text{and} \qquad \partial_k T_{ij}^{TT} = -\mu_G \mathscr{B}_{kij} \qquad (13)$$

These equations are directly analogous to the constituent equations in (7) for the electric and magnetic fields. For a sinusoidal stress tensor, we have $\partial_t T_{ij}^{TT} \propto \omega T_{ij}^{TT}$. Therefore, in the DC limit ($\omega \to 0$), the first equation in (13) requires that $\mathscr{E}_{ij} \to 0$. This implies that in the DC limit, the propagating electric-like tensor field vanishes completely throughout the entire superconductor, and only the propagating magnetic-like tensor field, $\mathscr{B}_{ijk}$, remains. It can also be shown that the wave equation in (2) and the definitions in (10), lead to

$$\partial_k \mathscr{B}_{kij} = -2\kappa T_{ij}^{TT} - \frac{1}{c^2}\partial_t \mathscr{E}_{ij} \qquad (14)$$

---

[4] It is shown in [4] that $h_{ij}^{TT}$ is the only part of the metric perturbation that satisfies a wave equation. All other components of the metric perturbation lead to Poisson equations and therefore can be neglected in the far field. Therefore we have isolated only the propagating degrees of freedom by choosing $h_{00} = h_{0i} = 0$ and $h_{ij} = h_{ij}^{TT}$.

[5] Similarly, the geodesic *deviation* equation for two test particles separated initially by a distance $L^i$ in the presence of a gravitational wave (to linear order in the metric and first order in velocity) can be written in terms of $\mathscr{E}_{ij}$ and $\mathscr{B}_{ijk}$ as

$$\ddot{L}^i = -\frac{1}{2}L^j \left[\dot{\mathscr{E}}_{ij} + v^k \left(\dot{\mathscr{B}}_{ikj} + \dot{\mathscr{B}}_{jki} - 2\dot{\mathscr{B}}_{ijk}\right)\right] \qquad (12)$$

where we have chosen the local Lorentz frame of one of the test particles.

This is essentially an Ampere-like law in the sense that a spatial derivative of a magnetic-like tensor field is proportional to a source term plus a time-derivative of the electric-like tensor field. For sinusoidal stress, the DC limit requires $\mathcal{E}_{ij} \to 0$. Taking a spatial derivative of (14) and inserting the second equation of (13), leads to

$$\nabla^2 \mathcal{B}_{ijk} - 2\kappa \mu_G \mathcal{B}_{ijk} = 0 \tag{15}$$

This is a Yukawa-like equation similar to (8) which implies an exponential decay solution for $\mathcal{B}_{ijk}$ given by $\mathcal{B}_{ijk}(z) = \mathcal{B}_{ijk(0)} e^{-z/\lambda_G}$, where $\mathcal{B}_{ijk(0)}$ is a constant amplitude tensor, $z$ is the distance from the surface to the interior of the superconductor, and $\lambda_G$ is a gravitational penetration depth in the DC limit. It follows that

$$\lambda_G = \frac{c^2}{\sqrt{16\pi G \mu_G}} = \frac{c}{\omega_G} \tag{16}$$

which is consistent with the DC limit of the gravitational penetration depth found in (6). This result implies that the magnetic-like tensor field is expelled from the superconductor in a gravitational Meissner-like effect.[6]

Therefore, we conclude that a Meissner-like effect leading to the expulsion of the gravitational wave field is valid for *all* frequencies down to the DC limit. For an upper bound on the frequencies, we would expect that the BCS energy gap frequency would limit the maximum frequency permitted for this Meissner-like effect to occur since frequencies above this would break up the Cooper pairs and destroy the superconducting state.

## 5 The Ginzburg-Landau free energy density embedded in curved space-time

In this section, we will formulate the response of the Cooper pair density to a gravitational wave using the Ginzburg-Landau (GL) free energy density embedded in curved space-time. First recall that the non-relativistic GL free energy density in flat space-time can be written as [7]

$$\mathscr{F} - \mathscr{F}_n = \frac{1}{2m} \left| \left( -i\hbar\nabla - q\mathbf{A} \right) \psi \right|^2 + \alpha |\psi|^2 + \frac{\beta}{2} |\psi|^4 \tag{17}$$

where $m = 2m_e$, $q = 2e$, $\alpha$ and $\beta$ are phenomenological parameters, and $\psi$ is a complex order parameter. (We exclude the contribution of the magnetic field energy density.) To obtain a GL free energy density with coupling to gravitational fields, we begin with the relativistic invariant, $g_{\mu\nu}\pi^\mu\pi^\nu = -m^2c^2$, where $\pi^\mu = (E/c, p^i)$ is the kinetic four-momentum. Expanding the metric as $g_{\mu\nu} = \eta_{\mu\nu} + h_{\mu\nu}$, solving for the energy, remaining to first order in the metric perturbation, and remaining to second order in velocity, gives

$$E = \left(1 - \frac{1}{2}h_{00}\right)mc^2 + \left(1 - \frac{1}{2}h_{00}\right)\frac{\pi^2}{2m}$$
$$+ ch_{0i}\pi^i + \frac{h_{ij}\pi^i\pi^j}{2m} \tag{18}$$

For gravitational waves in the far-field, we can reduce the metric perturbation to just the transverse-traceless part so that $h_{\mu 0} = 0$ and $h_{ij} = h_{ij}^{TT}$. Also, in the non-relativistic limit, it is standard to drop the rest energy, $mc^2$, giving[7]

$$E = \frac{\pi^2}{2m} + \frac{h_{ij}^{TT}\pi^i\pi^j}{2m} \tag{19}$$

We now find the Helmholtz free energy, $F = -k_B T \ln(Z)$, where the partition function for a canonical ensemble is given by $Z = \sum_n \exp(-\beta E_n)$, with $E_n$ being the energy modes of the system and $\beta = (k_B T)^{-1}$. For a superconductor, all the Cooper pairs are considered as condensed into a Bose-Einstein condensate where the particles are in the zero-momentum ground state and therefore behave as effectively a single coherent particle. In that case, the Helmholtz free energy is equivalent to the energy in (19).

To describe a superconductor, we can use a procedure similar to Ginzburg-Landau. We introduce a complex order parameter, $\psi$, as well as a quadratic potential, $\alpha|\psi|^2$, and a quartic self-coupling term, $\beta|\psi|^4$. Also, to formulate the free energy density in *curved* space-time[8], we use the coordinate volume expressed in terms of the proper volume as $dV = dV_{proper}/\sqrt{-g^{\tau\tau}}$, where $\sqrt{-g^{\tau\tau}}$ is the Jacobian and $g^{\tau\tau}$ is the determinant of the metric in terms of the transverse-traceless part of the metric perturbation. Using the notation $\mathscr{F}_{GL} = \mathscr{F} - \mathscr{F}_n$, we can write the GL free energy density with respect to proper volume as

$$\mathscr{F}_{GL} = \sqrt{-g^{\tau\tau}} \left[ \frac{1}{2m} |\pi^i \psi|^2 + \frac{1}{2m} h_{ij}^{\tau\tau} (\pi^i \psi)^* (\pi^j \psi) \right.$$
$$\left. + \alpha |\psi|^2 + \frac{\beta}{2} |\psi|^4 \right] \qquad (20)$$

where $\pi^i = -i\hbar\partial^i - qA^i$ for coupling to electromagnetism. If we define a quantum stress as $T^{ij} \equiv \frac{1}{2m}(\pi^i\psi)^*(\pi^j\psi)$, then we notice that the coupling of the gravitational wave field to the quantum stress of the superconductor has the form $h_{ij}^{\tau\tau} T^{ij}$ which is consistent with the coupling found in [10].

---

[6] Although there is a Meissner-like effect for the gravitational wave strain field in the DC limit, there is no corresponding Meissner-like effect for Earth's gravitational field. This is due to the fact that the gravitational field of the Earth can be shown to satisfy a Helmholtz-like equation rather than a Yukawa-like equation in a superconductor. Therefore, a superconductor does not permit the shielding of Earth's gravitational field. [6]

[7] Dropping $mc^2$ allows (20) to reduce to (17) in the absence of a gravitational wave as expected. In fact, $mc^2$ should not appear in the Helmholtz free energy (20) since the rest energy is not an accessible energy of the system (in this context).

[8] For a relativistic covariant form of the Ginzburg-Landau equation in *flat* space-time, refer to Section 2.1 of [8].

The GL complex order parameter can be written as $\psi = \sqrt{n_s} \exp\left(\frac{i}{\hbar}\mathbf{p}_0 \cdot \mathbf{r}\right)$, where $n_s$ is the number density of Cooper pairs, and $\mathbf{p}_0$ is the momentum eigenvector in the ground state. Since the Cooper pairs are in a zero-momentum eigenstate $(p_0 = 0)$, then $\psi^*\psi = n_s$ which is a constant. Therefore, when using $\pi^i = -i\hbar\partial^i - qA^i$ in (20), we find that all the derivatives vanish and we obtain

$$\mathscr{F}_{GL} = \sqrt{-g^{\tau\tau}} \left[ \frac{q^2 n_s}{2m}(A^i)^2 + \frac{q^2 n_s}{2m} h_{ij}^{\tau\tau} A^i A^j \right.$$
$$\left. + \alpha n_s + \frac{\beta n_s^2}{2} \right] \qquad (21)$$

Now to consider the work done on a system due to a gravitational strain, $h_{ij}$, causing a stress, $T_{ij}$, we use $\mathscr{W} = \int T^{ij} dh_{ij}$, where $\mathscr{W}$ is the work density (or work per unit volume). This means that $T^{ij} = d\mathscr{W}/dh_{ij}$. By the work-energy theorem, we recognize that work done by the system must reduce the internal energy density so that $\Delta \mathscr{U} = -\mathscr{W}$. We also use the thermodynamic relation $dU = TdS - PdV$ with the Helmholtz free energy being $F = U - TS$. When stresses and strains are present, then the Helmholtz free energy *density* satisfies the relation $d\mathscr{F} = -\mathbb{S}dT + T^{ij}dh_{ij}$, where $\mathbb{S}$ is the entropy density.

It then follows that the stress produced by a gravitational wave interacting with the Cooper pair density can be expressed in terms of the GL free energy density as

$$T^{ij} = \left(\frac{\partial \mathscr{F}_{GL}}{\partial h_{ij}}\right)_T \qquad (22)$$

For a gravitational wave propagating in the $z$-direction, $h_{i3}^{\tau\tau} = 0$ (since $h_{ij}^{\tau\tau}$ is transverse). For a plus polarization wave, we have $h_{11}^{\tau\tau} = -h_{22}^{\tau\tau} = h_\oplus$ which corresponds to a stress $T_{11}^{\tau\tau} = -T_{22}^{\tau\tau} = T_\oplus$. For a cross polarization wave, we also have $h_{12}^{\tau\tau} = h_{21}^{\tau\tau} = h_\otimes$ which corresponds to a stress $T_{12}^{\tau\tau} = T_{21}^{\tau\tau} = T_\otimes$. Therefore, we can write (22) as

$$T_\oplus = \left(\frac{\partial \mathscr{F}_{GL}}{\partial h_\oplus}\right)_T \quad \text{and} \quad T_\otimes = \left(\frac{\partial \mathscr{F}_{GL}}{\partial h_\otimes}\right)_T \qquad (23)$$

To obtain expressions for $T_\oplus$ and $T_\otimes$ that are purely in terms of $h_\oplus$ and $h_\otimes$, respectively, we can consider the cases of plus and cross polarization waves independently. Then writing (1) as $T_\oplus = -\mu_G h_\oplus$ and $T_\otimes = -\mu_G h_\otimes$ for the two polarizations, we find that the gravitational shear modulus for either polarization becomes

$$\mu_{G(CP)} = \frac{e^2 n_s}{m_e}(A^i)^2 + \alpha n_s + \frac{\beta}{2} n_s^2 \qquad (24)$$

where the subscript "CP" represents Cooper pairs. For the zero-momentum eigenstate, the minimal coupling rule reduces to $mv^i = -qA^i$. Therefore, the first term in (24) becomes $\frac{1}{2} m n_s (v^i)^2$ which is the kinetic energy density of the Cooper pair supercurrent. We can also combine the last two terms of (24) by using the coherence length, $\xi = \hbar/\sqrt{2m_e|\alpha|}$, and the relation $n_s = -\alpha/\beta$ which comes from minimizing (17). Then (24) becomes

$$\mu_{G(CP)} = \frac{m n_s}{2}(v^i)^2 + \frac{\hbar^2 n_s}{4m_e \xi^2} \qquad (25)$$

The second term characterizes the minimum response of the Cooper pair density to a gravitational wave in the absence of supercurrents. Since $\xi$ is a phenomenological parameter describing the superconductor, it follows that this term also becomes a phenomenological parameter - an intrinsic property of the material.

To obtain a numerical result, we consider that each atom contributes two conduction electrons, and only $10^{-3}$ of the conduction electrons are in a superconducting state [7], so that $n_s \approx 2n(10^{-3})$, where $n = \rho/m$ is the number density of atoms. For niobium, the mass density is $\rho \approx 8.6 \times 10^3 \, kg/m^3$ and the mass per atom is $m \approx 1.5 \times 10^{-25} \, kg/atom$. Then the number density of atoms is $n \approx 5.7 \times 10^{28} \, m^{-3}$ and therefore the number density of Cooper pairs is $n_s \approx 1.1 \times 10^{26} \, m^{-3}$. Using $\xi \approx 39 \, nm$ for the coherence length of niobium makes the second term of (25) approximately $2.2 \times 10^2 \, J/m^3$.

For an upper limit on the first term, the maximum kinetic energy which will preserve the superconducting state is determined by the BCS energy gap, $E_{gap} = \frac{7}{2} k_B T_c$, where $k_B$ is the Boltzmann constant and $T_c$ is the critical temperature. For niobium, $T_c = 9.3 K$, so the energy density of the BCS energy gap is $E_{gap} n_s \approx 4.9 \times 10^4 \, J/m^3$. Since this term dominates the second term in (25), then we have

$$\mu_{G(CP)} \approx 4.9 \times 10^4 \, J/m^3 \qquad (26)$$

To determine the *material* strain that would occur for the Cooper pair density, we can use the corresponding *material* constituent equation, $T_{ij}^{\tau\tau} = -s u_{ij}^{\tau\tau}$, where $s$ is the *material* shear modulus and $u_{ij}^{\tau\tau}$ is the dimensionless strain tensor of the material. By equating this *material* constituent equation with the *gravitational* constituent equation, $T_{ij}^{\tau\tau} = -\mu_G h_{ij}^{\tau\tau}$, the material strain can be expressed in terms of the gravitational strain as

$$u_{ij}^{\tau\tau} = \frac{\mu_G}{s} h_{ij}^{\tau\tau} \qquad (27)$$

For niobium, the material shear modulus is known to be $s \approx 3.8 \times 10^{10} J/m^3$. Therefore, (26) predicts that the material strain is related to the gravitational strain for the Cooper density by a factor of $\frac{\mu_{G(CP)}}{s} \approx 10^{-6}$. If the amplitude of the gravitational wave is on the order of $h_{ij}^{\tau\tau} \sim 10^{-20}$, then the material strain produced would be on the order of $u_{ij}^{\tau\tau} \sim 10^{-26}$. This demonstrates that the material strain is tiny for the Cooper pair density in response to the gravitational wave.

## 6 The Debye free energy in the low-temperature limit embedded in curved space-time

We now describe the lattice ions using the Debye model in the low-temperature limit with coupling to gravitational waves in curved space-time. Using the approach of [11], we find that in the non-relativistic limit, for gravitational waves in the far-field, the Hamiltonian is

$$\hat{H} = \frac{\hat{p}^2}{2m} + \frac{h_{ij}^{\tau\tau} \hat{p}^i \hat{p}^j}{2m} + \frac{K}{4} \hat{x}_i^2 \qquad (28)$$

where we have included a harmonic potential term to model the lattice ions as quantum harmonic oscillators.

For the case of a *plus* polarization gravitational wave, we set $h_\otimes = 0$. Then we act the Hamiltonian on a state, $\Psi(\mathbf{x}, t)$, and use the quantum operator for the Hamiltonian, $\hat{H} = i\hbar\partial_t$, and canonical momentum, $\hat{p}_i = -i\hbar\partial_i$. We also consider a separable solution which can be written as $\Psi(\mathbf{x}, t) = \psi(\mathbf{x})\varphi(t)$. This leads to

$$\frac{i\hbar}{\varphi(t)}\partial_t \varphi(t) = -\frac{\hbar^2}{2m}\frac{1}{\psi(\mathbf{x})}\left[(1+h_\oplus)\partial_x^2 + (1-h_\oplus)\partial_y^2\right]\psi(\mathbf{x})$$
$$-\frac{\hbar^2}{2m}\frac{1}{\psi(\mathbf{x})}\partial_z^2 \psi(\mathbf{x}) + \frac{K_i}{4}\hat{x}_i^2 \qquad (29)$$

We equate the left side to $E(t)$ which is the energy as a function of time. Then we have $\partial_t \varphi(t) = -\frac{iE(t)}{\hbar}\varphi(t)$. If we define $\varepsilon(t) \equiv \int_0^t E(t')\,dt'$, then the solution to the differential equation has the form $\varphi(t) = e^{-i\varepsilon(t)t/\hbar}$. Hence we find that the gravitational wave will introduce a phase shift to the quantum wave function which is determined by $\varepsilon(t)$. Returning to (29), we can equate the right side to $E(t)$ and multiply through by $\psi(\mathbf{x})$.

$$E(t)\psi(\mathbf{x}) = -\frac{\hbar^2}{2m}\left[(1+h_\oplus)\partial_x^2 + (1-h_\otimes)\partial_y^2\right]\psi(\mathbf{x})$$
$$-\frac{\hbar^2}{2m}\partial_z^2\psi(\mathbf{x}) + \frac{K_i}{4}\hat{x}_i^2\psi(\mathbf{x}) \qquad (30)$$

Notice that this is *not* a time-*independent* Schrödinger equation. The states are time-independent, but the coefficients, $h_\oplus(z, t)$ and $h_\otimes(z, t)$, make the Hamiltonian time-*dependent*. We can separate the time-dependent terms of the Hamiltonian by writing the expression above as

$$E(t)\psi(\mathbf{x}) = \left[-\frac{\hbar^2}{2m}\left(\partial_x^2 + \partial_y^2 + \partial_z^2\right) + \frac{K_i}{4}\hat{x}_i^2\right]\psi(\mathbf{x})$$
$$-\frac{\hbar^2}{2m}\left[h_\oplus(z,t)\partial_x^2 - h_\otimes(z,t)\partial_y^2\right]\psi(\mathbf{x}) \qquad (31)$$

The first line can now be considered as an *unperturbed* Hamiltonian, $\hat{H}_0$, while the second line is a time-*dependent* perturbation, $\hat{H}_1(t)$. For a gravitational wave with *periodic* time-dependence, the full Hamiltonian has a periodic behavior and thus applying Floquet's theorem leads to quasi-energy eigenvalues.[9] Using a standing gravitational wave in the $z$-direction given by $h_\oplus(z,t) = A_\oplus \cos(kz)\sin(\omega t)$, and using a thin film approximation, $z << \lambda$, we obtain the following quasi-energy eigenvalues.

$$E = \frac{1}{2m}\left(p_x^2 + p_y^2 + p_z^2\right) + \frac{K_i}{4}x_i^2$$
$$+ \frac{1}{4\pi m}A_\oplus\left(p_x^2 - p_y^2\right) \qquad (32)$$

We now use $(\hat{x}_i)^2 = \hat{x}^2 + \hat{y}^2 + \hat{z}^2$ and $K_x$, $K_y$ and $K_z$ to characterize the harmonic potential in the $x$, $y$, and $z$-directions, respectively. Then we use $n_x$, $n_y$, and $n_z$ to denote the occupation numbers of the modes with frequencies $\omega_x$, $\omega_y$, and $\omega_z$ in the three directions, respectively. Applying separation of variables and summing over $N$ oscillators, with $n_\alpha$ being the number of phonons with frequency $\omega_\alpha$, we obtain the following quasi-energy eigen-values in terms of phonon modes.

---

[9] The method is described by Zel'dovich [12] and later with further detail by Sambe [13].

$$E_{(n_x,n_y,n_z)} = \hbar \sum_\alpha^N \omega_\alpha \left[ \sqrt{A_\oplus^+} \left(n_{x,\alpha} + \tfrac{1}{2}\right) \right.$$
$$\left. + \sqrt{A_\oplus^-} \left(n_{y,\alpha} + \tfrac{1}{2}\right) + \left(n_{z,\alpha} + \tfrac{1}{2}\right) \right] \quad (33)$$

where $A_\oplus^\pm \equiv 1 \pm \frac{A_\oplus}{2\pi}$. Note that the coupling to a gravitational wave breaks the usual spatial isotropy of the modes in each spatial direction. This includes the zero-point energy of each oscillator, which implies that the gravitational wave also breaks the spatial isotropy of the *vacuum*. Since the gravitational wave is propagating in the $z$-direction, it modulates the frequencies in the $x$ and $y$ directions due to the squeezing/stretching of *space itself*. The factors, $\sqrt{A_\oplus^+}$ and $\sqrt{A_\oplus^-}$, can be considered as "gravitational modulation factors" which determine the modulation of the frequencies (and corresponding energies) for the quantum harmonic oscillators. The gravitational wave effectively shortens/lengthens the boundary conditions of the lattice and therefore dynamically modulates the vacuum energy. (This is analogous to the *mechanical* oscillation of conducting plates which leads to the standard electromagnetic dynamical Casimir effect.) Since the zero-point energy is that of lattice *phonons*, then this is a "dynamical gravito-phonon Casimir effect."

The physical meaning of this dynamical Casimir effect is that an increase in the occupation number of the phonon modes of the lattice is predicted to occur in the presence of the gravitational wave. We interpret this effect as a quantum-mechanical analog of a "Weber-bar effect". That is to say, the amplitude of the sound waves in the lattice grows in the classical limit due to the coupling of energy from the gravitational wave into energy in the modes of the lattice. Thus gravitational wave energy in the vacuum which is incident upon the superconductor could be converted into sound wave energy in the lattice, just like in the Weber bar.

We now evaluate the free energy using the energy eigenvalues in (33). The partition function for this canonical ensemble is[10]

$$Z = Z_0 \sum_n \exp\left\{ -\beta\hbar \sum_\alpha^N \omega_\alpha \left[ \sqrt{A_\oplus^+} \left(n_{x,\alpha} + \tfrac{1}{2}\right) \right. \right.$$
$$\left. \left. + \sqrt{A_\oplus^-} \left(n_{y,\alpha} + \tfrac{1}{2}\right) + \omega_{z,\alpha}\left(n_{z,\alpha} + \tfrac{1}{2}\right) \right] \right\} \quad (34)$$

---

[10] Here we are essentially applying the concept of "periodic thermodynamics" as described by Kohn in [14] and more recently by Langemeyer and Holthaus in [15].

where $Z_0 = e^{-\beta E_0}$ and $E_0 = \tfrac{1}{2}\hbar N \omega \left(\sqrt{A_\oplus^+} + \sqrt{A_\oplus^-} + 1\right)$ is the zero-point energy. Next, we find the Helmholtz free energy, $F = -k_B T \ln(Z)$, and apply the Debye model in the low temperature limit. This leads to the following free energy density embedded in curved space-time.

$$\mathscr{F}_D = \sqrt{-g_\oplus^{\tau\tau}} \left[ \frac{1}{2} n\hbar\omega \left(\sqrt{A_\oplus^+} + \sqrt{A_\oplus^-} + 1\right) \right.$$
$$\left. - \frac{\pi}{6\hbar\beta^2 v} \left( \frac{1}{L_y L_z \sqrt{A_\oplus^+}} + \frac{1}{L_x L_z \sqrt{A_\oplus^-}} + \frac{1}{L_x L_y} \right) \right] \quad (35)$$

where $L_x$, $L_y$, $L_z$ are the dimensions of the superconductor, and $g_\oplus^{\tau\tau}$ is the determinant of the metric in terms of the transverse-traceless metric perturbation for a plus polarization gravitational wave. Similar to (22), we write the stress produced by a gravitational wave interacting with lattice ions in terms of the Debye free energy density as

$$T_\oplus = \left(\frac{\partial \mathscr{F}_D}{\partial A_\oplus}\right)_T \quad (36)$$

Likewise, (1) becomes $T_\oplus = -\mu_{G(LI)} A_\oplus$ where the subscript "LI" represents the lattice ions. Expanding (35) to second order in the metric perturbation, applying (36), and assuming $L_x \approx L_y \gg L_z$ for a thin superconducting film, leads to the following gravitational shear modulus for the lattice ions.

$$\mu_{G(LI)} \approx \frac{3\hbar\omega n}{2} - \frac{\pi}{3\hbar\beta^2 v L_x L_z} \quad (37)$$

The first term can be referred to as the "zero-point energy density" term since it originates from the contribution of the zero-point energy to the free energy of the system. For an upper bound on this value, we can use the Debye frequency, $\omega_D$, which is the cut-off frequency in the Debye model. For a niobium superconductor, $\omega_D \approx 3.1 \times 10^{13}\, s^{-1}$. The number density of atoms was also found earlier to be $n \approx 5.7 \times 10^{28}\, m^{-3}$. These lead to a value of $2.8 \times 10^8\, J/m^3$ for the "zero-point energy density" term in (37).

The second term in (37) can be referred to as the "sum of modes" term since it originates from the contribution of the sum of all the other modes of the system. To obtain a value for this term, we can use a thickness for the superconducting film on the order of micrometers $(L_z \approx 10^{-6}\, m)$ and a surface with edges on the order of centimeters $(L_x \approx 10^{-2}\, m)$. The velocity is $v = \sqrt{s/\rho}$, where $\rho = 8.6 \times 10^3\, kg/m^3$ is the mass density of the ionic lattice and $s \approx 3.8 \times 10^{10}\, J/m^3$ is the material shear modulus for niobium. Then $v \approx 2.1 \times 10^3\, m/s$. We can also use $T \approx 10^{-2}\, K$ for the temperature of the superconductor. This leads to a value of $9.0 \times 10^{-12}\, J/m^3$ for the "sum of modes" term. Therefore, we find that the zero-point energy density term clearly dominates over the "sum of modes" term and the gravitational shear modulus for the ionic lattice of niobium can be considered to have a value given by

$$\mu_{G(LI)} \approx 2.8 \times 10^8 \, J/m^3 \tag{38}$$

Comparing this to (26) for the Cooper pair density, we find that the lattice ions are on the order of $10^4$ times more responsive to a gravitational wave than the Cooper pairs.

## 7 Conclusion

The relative strain between the ionic lattice and the Cooper pair density is the difference between $u^{\tau\tau}_{ij(CP)}$ for the Cooper pairs and $u^{\tau\tau}_{ij(LI)}$ for the ionic lattice. Therefore, we can write the relative strain induced by the gravitational wave as $U^{\tau\tau}_{ij} \equiv u^{\tau\tau}_{ij(LI)} - u^{\tau\tau}_{ij(CP)}$. Using (27) gives

$$U^{\tau\tau}_{ij} = \frac{\mu_{G(LI)} - \mu_{G(CP)}}{s} h^{\tau\tau}_{ij} \tag{39}$$

Inserting the results from (26) and (38), and making use of $s \approx 3.8 \times 10^{10} \, J/m^3$ for niobium, we find that (39) gives

$$U^{\tau\tau}_{ij} \approx \left(10^{-2}\right) h^{\tau\tau}_{ij} \tag{40}$$

This result demonstrates that there is a relative strain between the ionic lattice and the Cooper pair density that is approximately one percent of the gravitational wave strain. Since lattice ions are positively charged and Cooper pairs are negatively charged, then we find that a charge separation effect will occur in the superconductor in response to a gravitational wave. Note that the difference between $\mu_{G(CP)}$ and $\mu_{G(LI)}$ in (39) is due to quantum mechanical phenomena (namely, the zero-point energy of the ionic lattice and the BCS energy gap of the superconductor), therefore this charge separation effect is a macroscopic *quantum* effect which has no classical analog.

Lastly, we note that the charge separation effect is expected to occur only within the London penetration of the superconductor, where non-zero electromagnetic fields can exist. In fact, due to the gravitational Meissner-like effect, the gravitational wave would exponentially decay within the penetration depth of the superconductor so that the charge separation effect could not be induced in the interior regions of the superconductor.

## 8 Acknowledgements

The authors thank G. Muñoz and D. Singleton for helpful discussions.

**Key words.** General Relativity, Gravitational waves, Ginzburg-Landau theory of superconductivity, Debye model, Casimir effect

## References

[1] S. Minter, K. Wegter-McNelly, and R. Chiao, "Do mirrors for gravitational waves exist?", Physica E, 42, 234 (2010); arXiv:0903.0661.
[2] Q. Quach, "Gravitational Casimir Effect," Phys. Rev. Lett. 114, 081104 (2015).
[3] C. Misner, K. Thorne, and J. Wheeler, *Gravitation* (San Francisco: Freeman, 1972).
[4] É. Flanagan and S. Hughes, "The basics of gravitational wave theory," New J. Phys. 7, 204 (2005). We are primarily concerned with Section 2.2, "Global spacetimes with matter sources," pp. 9-15.
[5] W. Press, "On Gravitational Conductors, Waveguides, and Circuits," General Relativity and Gravitation, Vol. 11, No. 2 (1979).
[6] H.A. Chan, M.V. Moody, H.J. Paik, "Superconducting gravity gradiometer for sensitive gravity measurements. II. Experiment," Phys. Rev. D 35, 3572 (1987).
[7] M. Tinkham, *Introduction to Superconductivity*, 2nd edition (McGraw Hill, Inc., 1976).
[8] D. Bertrand, "A Relativistic BCS Theory of Superconductivity," Dissertation, Université catholique de Louvain, (2005).
[9] O. Dinariev, B. Mosolov, "Relativistic Generalization of the Ginzburg-Landau Theory," UDC 530.12+537.312.62 (1986), translated from Izvestiya Vysshikh Uchebnykh Zavedenii, Fizika, No. 4, 98 (1989).
[10] T. Rothman, S. Boughn, "Aspects of Graviton Detection: Graviton Emission and Absorption by Atomic Hydrogen," Found. Phys. 36, 1801 (2006).
[11] B. DeWitt, "Superconductors and gravitational drag," Phys. Rev. Lett. 16, 1092 (1966); G. Cognola, et al., "Relativistic wave mechanics of spinless particles in a curved space-time," General Relativity and Gravitation, 18, 971 (1986).
[12] Y. Zel'dovich, "The Quasienergy of a Quantum-Mechanical System Subjected to a Periodic Action," Soviet Physics JETP, Vol. 24, Number 5 (1967).
[13] H. Sambe, "Steady States and Quasienergies of a Quantum-Mechanical System in an Oscillating Field," Phys. Rev. A 7, 6 pp. 2203-2213 (1973).
[14] W. Kohn, "Periodic Thermodynamics," Journal of Statistical Physics, Vol. 103, Issue 3-4, 417 (2001).
[15] M. Langemeyer, M. Holthaus, "Energy flow in periodic thermodynamics," Phys. Rev. E, 89 (2014).